\RequirePackage{fix-cm}

\documentclass[smallextended]{svjour3VISbi}       
\smartqed  
\usepackage{graphicx}
\usepackage{color}

\usepackage{amssymb}
 \journalname{Quantum Studies: Mathematics and Foundations (2017)  1-19}
 \doi{DOI: 10.1007/s40509-017-0116-z}
\newcommand{\Schrodinger}{Schr\"{o}dinger~}
\begin{document}

\bibliographystyle{spmpsci}      

\title{Hydrodynamics of Superfluid Quantum Space: \\ de Broglie interpretation of the quantum mechanics}


\titlerunning{Hydrodynamics of Superfluid Quantum Space: de Broglie interpretation ...}        

\author{Valeriy I. Sbitnev}


\institute{V. I. Sbitnev \at
              St. Petersburg B. P. Konstantinov Nuclear Physics Institute, NRC Kurchatov Institute, Gatchina, Leningrad district, 188350, Russia; \\
               Department of Electrical Engineering and Computer Sciences, University of California at Berkeley, Berkeley, CA 94720, USA\\ \\            
              \email{valery.sbitnev@gmail.com}      
}

\date{\today}

\maketitle

\begin{abstract}
The ubiquitous ether coming from the ancient times up to middle of the twenty century is replaced by a superfluid quantum space. It represents by itself a Bose-Einstein condensate consisting of enormous amount of virtual particle-antiparticle pairs emerging and disappearing in an infinitely ongoing dance. Flowing of this medium in the non-relativistic limit  is described by the modified Navier-Stokes equation along with the continuity equation. The first equation admits the splitting on to two coupled equations. They are the quantum Hamilton-Jacobi equation and the equation for vorticity. The quantum Hamilton-Jacoby equation paired with the continuity equation can be reduced to the \Schrodinger equation. These two equations representing the kernel of the Bohmian mechanics give finding bundle of the Bohmian trajectories. Whereas the vorticity equation gives solutions for vortices moving along such trajectories. As the result we come to the de Broglie's interpretation of quantum mechanics according to which there is a pilot-wave guiding the particle (in our case it is a vortex clot) from a source up to its detection along an optimal path  that is the Bohmian trajectory.


\end{abstract}

\section{\label{sec1}Introduction}

The issue concerning what space, time and motion are accompanies the humanity all along since antiquity. The most distinct form of representation about space and time has developed in the form of two dialectically opposite ideas~\cite{Stanford2015}, later known as the Democritus-Newton and Aristotle-Leibniz conceptions. According to Democritus every thing is made up of tiny, invisible, indivisible corpuscles, named "atoms". Atoms diffuse in the space or combine.  Between atoms we have empty space. On the other hand, Aristotle, student of Plato's Academy in Athens, believed that Universe is presented in two parts, the terrestrial region and the celestial (over the moon). The terrestrial region is comprised of combinations of four substances, earth, fire, air, and water. Whereas the celestial region (Sun, planets, stars) consists of  ether, a special kind of matter that is not found on Earth. It is the all-pervading ether filling all space, therefore emptiness is not exist.

The ether, named also by fifth element, possesses cardinally other qualities than the four above mentioned terrestrial elements. 
Unusual properties of the ether attracted the attention of naturalists for explaining propagation of the light through our 3D space~\cite{Whittaker1990}.  Christian Huygens, a contemporary of Newton, believed that the space is everywhere densely filled with an extraordinary medium which he also calls ether~\cite{Huygens1912}, through which light may spread from point to point due to a secondary re-emission of these elements of ether. 

The understanding of the mechanisms  of light propagation through the space started with James Maxwell's equations of the electromagnetic field. However, the question arises how electric and magnetic fields are transmitted over large distances through the empty space. Naturally to assume that a medium supporting the transmission of the electromagnetic fields at distances is the ether. However, Maxwell's attempt to build a model of such a medium which would be like liquid, gas, or solid crystal, was not successful.  Perhaps the failure was that this model had been built in accordance with the hallmarks of classical mechanics. 
The classical ether which could decelerate or accelerate light depending on oncoming or accompanying the earth's rotation (ether dragging)  was rejected after the experiment of Michelson-Morley which was specifically performed for that aim.

Although the ether was expelled from physics, the attempts to rehabilitate it did not stop until now~\cite{deBroglie1987,Dirac1951,PetroniVigier1983,SinhaEtAl1976,Whittaker1990}.
But nowadays, instead of the word ether physicists use a more neutral combination "quantum vacuum". In favor of this rehabilitation, there are serious experimental and theoretical grounds. A characteristic and refined example, obtained at the tip of the pen, is the energy of the quantum oscillator~\cite{t_Hooft2014}
\begin{equation}
    E_{n} = \hbar\omega\Bigl(
                                   n + {{1}\over{2}}
                                   \Bigr),
   ~~~~ n = 0, 1, 2, \cdots.                                   
\label{eq=1}
\end{equation}
 One can see that the energy of the medium, entirely free from the field quanta, $n=0$, has the nonzero value 
 $E_{0} = \hbar\omega/2$. This residual level is due to zero-point vacuum fluctuations.

From this point we are immersed in the quantum realm.
One of the major mysteries of quantum mechanics remains that of the emergence of the \Schrodinger equation~\cite{Schrodinger1926}. This equation formulated by Erwin \Schrodinger in 1925 and  published in 1926 plays in quantum mechanics the same important role as Hamilton equations or the Newton's second law in classical mechanics or Maxwell's equations for electromagnetic waves. 
After about 100 years dealing with the \Schrodinger equation its  solutions repeatedly confirmed a good agreement with experiments. So we may adopt this equation as a reference of the non-relativistic quantum mechanics.

Almost simultaneously with the \Schrodinger article the Madelung article entitled "Quantum theory in hydrodynamical form"~\cite{Madelung1926} appeared in print. Madelung had shown that the  \Schrodinger  equation  for  one-electron  problem  can  be  transformed  into  the  form of  hydrodynamical equations as soon as the wave function $\psi$ is written in the polar form, $\psi=\alpha\exp\{{\bf i}\beta\}$, where $\alpha=\sqrt{\rho}$ is the amplitude factor ($\rho$ is the density distribution) and $\beta$ is the wave phase. Further, a consistent separation of the solution on real and imaginary parts should be executed. As a result we get two equations. One of them is the Euler equation loaded by an extra quantum correcting term
\begin{equation}
  Q = -{{\hbar^2}\over{2m}}{{\Delta\alpha}\over{\alpha}}
\label{eq=2}
\end{equation}
named further as  the Bohm quantum potential~\cite{Bohm1952a,Bohm1952b}. Here $m$ is the electron mass and $\hbar$ is the reduced Planck constant.
The second equation is the continuity equation.  The Madelung equations can be viewed as an alternative description to the \Schrodinger equation by associating the behavior of a non-relativistic motion of the Madelung irrotational fluid~\cite{Tsekov2012,HeifetzCohen2015}. 

The hydrodynamic analogy can give important insights with regard to a quantum particle behavior described by the \Schrodinger equation. For that reason this approach is gaining wider interest of 
the scientific community~\cite{HeifetzCohen2015,RecamiSales1998,Sonego1991,Takabayasi1983}.
Note, however, that the Madelung's ideas differ from the de Broglie-Bohm program~\cite{Oriols2012,OriolsMompart2013}. The former regards an electron with mass $m$ and wave function $\psi({\vec r}; t)$ not as a particle with a determined trajectory but as a continuous fluid with mass density $m|\psi({\vec r}; t)|^2$~\cite{Takabayasi1983,Wyatt2005}, whereas the latter aims to predict what happens, in principle, to a single electron~\cite{BensenyEtAl2014,BohmVigier1954,Sanz2015}. 

Jean-Paul Vigier wrote in connection of that approach in his article~\cite{Vigier1982}: "we have assumed that elements of the fluid, as well as the particles, which we represent in the form of well localized inhomogeneities moving in general along the streamlines with their local four-velocity ${\vec{\mathit v}}(x_{\mu})$  (neglecting stochastic fluctuations), are now considered in a first approximation, as very small bilocal structures." And ibid: "another novelty is the assumption that used in the Bohm-Vigier theory~\cite{BohmVigier1954} elements of the fluid are composed of the above-mentioned oscillators. So our particles are the oscillators that move along the streamlines of the fluid, consisting of the oscillators, which are in phase with all the surrounding oscillators."

Special interest has Nelson's ideas that they are based on conceiving quantum mechanics as a stochastic process~\cite{PenaEtAl2015,Nelson1967,Nelson1985}. The hypothesis is that a free particle in empty space, or let us say the ether, is subject to Brownian motion with the diffusion coefficient $\nu$  inversely proportional to the particle mass, $\nu = \hbar/2m$, motion of which is adequately described by the Wiener process. A result of such a consideration is getting the \Schrodinger equation~\cite{Nelson1966}.

At present modern physics has moved far ahead from understanding the sub-quantum background as a simple Madelung fluid.
Modern physical representations show that the world we perceive with our senses (possibly, enhanced by physical devices), contains only 5\% of matter, called baryon matter. 95\% is inaccessible for our perception. This inaccessible essence is called dark energy and dark matter~\cite{Ardey2006,Fiscaletti2016,Huang2013,Huang2016}.
There is a guess that this 95\% of the dark substance consists of fluctuating  pairs of particles and antiparticles that are in an endless dance of creation and annihilation of these rotating players. This omnipresent ocean of dark substance is the Bose-Einstein condensate consisting of such fluctuating paired particles with integer total
 spin~\cite{BoehmerHarko2007,DasBhaduri2015,NishiyamaEtAl2004,Sbitnev2015c,SbitnevFedi2017}. Zero-point fluctuations of the physical vacuum represent merely topological defects on the 4D hyper-surface of this ocean.
Feynman's path integral technique~\cite{FeynmanHibbs1965} opens a possibility to calculate the Brownian-like trajectories of non-relativistic particles in such a fluctuating space~\cite{Comisar1965}. 
The point is that the relativistic vacuum fluctuations providing a deep probing of the environment give the contribution to formation of the constructive and destructive interference fringes paving the path forward for the real non-relativistic particle~\cite{Sbitnev2013a}. 
In fact, this mechanism is the same preparing de Broglie pilot-wave guiding the particle along the optimal path~\cite{deBroglie1987}.

In this article we describe the superfluid motion of the quantum space in the non-relativistic approximation. This problem is solved by involving the Navier-Stokes equation and the continuity equation. The original Navier-Stokes equation is known to describe the motion of a classical viscous fluid. Therefore the challenge lies in the modification of this equation so that one could describe the superfluid motion of the quantum space. It is solved in~Sec.\ref{sec2}. Here we solve the problem of appearance of the quantum potential and the problem of fluctuating viscosity. As a result we come to the \Schrodinger equation.
In Sec.~\ref{sec3} we study vortex solutions of the Navier-Stokes equation. We apply the topological transformations in order to study the emergence of the toroidal motions, helicoidal rings on the torus, and emergence of a vortex ball (vortex clot) as a mathematical pattern of a particle. Sec.~\ref{sec4} summarizes the results through the prism of the unresolved problems of interpretation of the quantum mechanics formulated by Vitaly Ginzburg in his Nobel prize speech~\cite{Ginzburg2007a}. 
Special attention here is paid to consideration of the early ideas of Louis de Broglie~\cite{deBroglie1960,deBroglie1987}, which were further developed by David Bohm~\cite{Bohm1952a,Bohm1952b,BohmVigier1954}.

\section{\label{sec2}The modified Navier-Stokes equation}

We begin from the modified Navier-Stokes equation~\cite{Sbitnev2015b,Sbitnev2016b}:
\begin{equation}
 \rho_{M} \biggl(
 {{\partial {\vec {\mathit v}}}\over{\partial\,t}}
 + ({\vec {\mathit v}}\cdot\nabla){\vec {\mathit v}}
       \biggr) 
  =      {\vec f({\vec r},t)}    
   \;-\;
   \underbrace{ \rho_{M} \nabla (P/\rho_{M})
 \; +\; \mu(t)\,\nabla^{\,2}{\vec {\mathit v}} }.
\label{eq=3}
\end{equation}
 Here $\rho_{M} = M/{\Delta V}$ is the mass density distributed in the volume ${\Delta V}$, and ${\vec f}({\vec r},t)$ is the external force per the volume ${\Delta V}$.
  The modifications  relate to  the last two terms embraced by the brace:\\
  (a) the pressure gradient is rewritten as
\begin{equation}
  \nabla P \rightarrow \rho_{M}\nabla P/\rho_{M} =
  \nabla P - P \nabla \ln(\rho_{M});
\label{eq=4}
\end{equation} 
 (b) the dynamical viscosity coefficient, $\mu$, depends by $t$  in such a manner that
\begin{equation}
 \langle \mu(t) \rangle = 0_{+}, ~~~~~
 \langle \mu(t)\mu(0) \rangle >0.
\label{eq=5}
\end{equation} 

  The first modification leads to the appearance of the quantum potential~(\ref{eq=2}). As a consequence, the pair of equations, the modified Navier-Stokes equation~(\ref{eq=3}) and the continuity equation
 \begin{equation}
   {{\partial \rho_{M}}\over{\partial\,t}} 
   + (\nabla\cdot{\vec{\mathit v}})\rho_{M} = 0,
\label{eq=6}
\end{equation}
 can be reduced to one equation for a complex-valued function, the wave function $\psi$, to the \Schrodinger equation~\cite{Sbitnev2016b}. The second modification opens a path leading to getting long-lived vortices whose cores possess non-zero radius~\cite{Sbitnev2016c}. They are templates of particles.
 
\subsection{\label{sec21}Quantum potential} 

 We begin from the modified pressure gradient
\begin{equation}
\label{eq=7}
  \rho_{{M}}\nabla\biggl(
                   {{P}\over{\rho_{{M}}}}
                   \biggr)
=   \rho\nabla\biggl(
                   {{P}\over{\rho}}
                   \biggr)
 = \nabla P - P\nabla\ln(\rho_{_{}}).
\end{equation} 
The mass density distribution, $\rho_{M}$, is linked with the density distribution, $\rho$, by the following simple formula
\begin{equation}
 \rho_{_{M}} = {{M_{}}\over{\Delta V}}={{m N}\over{\Delta V}} = m\rho.
\label{eq=8}
\end{equation}
 So, there are $N$ carriers within the unit volume $\Delta V$, all with the elementary mass $m$, which give a bulk mass $M$ of the fluid~\cite{JackiwEtAl2004}. The density distribution $\rho({\vec r},t)$ gives, in fact, the probability
  $\rho({\vec r},t)\Delta {\vec r}\Delta t$ of discovering a carrier near the point ${\vec r}$ at the time moment $t$.
 
 Returning to Eq.~(\ref{eq=7}) we note, that  only the first term, the pressure gradient $\nabla P$, is represented in the original Navier-Stoke  equation~\cite{LandauLifshitz1987,KunduCohen2002}. The second term, $P\nabla\ln(\rho_{_{}})$, is an extra term  describing change in the logarithm of the density distribution $\rho$ on the infinitesimal increment of length multiplied by $P$.  
From here it follows that the modified pressure gradient is conditioned by the pressure gradient, as a classical variation of the pressure, and by adding the gradient of the entropy, $\ln(\rho)$~\cite{Sbitnev2009}, multiplied by the pressure. The latter may be interpreted as an increment of the information flow per length multiplied by the pressure, what introduces a pure quantum effect. 
Let us show this.
With this aim first  we suppose that the pressure $P$  
can be rewritten as
the sum of two pressures $P_1$ and $P_2$. 

  As for the pressure $P_1$ we begin with Fick's law
 which says that the diffusion flux, $\vec{\mathit J}$, is proportional to the negative value of the mass density gradient, 
 ${\vec{\mathit J}} =-D\nabla\rho_{{M}} $~\cite{Grossing2010}. Here, $D$ is the diffusion coefficient.
 Since ${D}\nabla{\vec{\mathit J}}$ has dimension of the pressure, we define the first pressure:
\begin{equation}
\label{eq=9}
  P_1 = {D}\nabla{{\vec{\mathit J}}}
 = -D^{2}\nabla^{2}\rho_{_{M}}.
\end{equation} 
  Observe that the kinetic energy of the diffusion flux of the fluid medium is $(M/2)({\vec{\mathit J}}/\rho_{_{M}})^{2}$.
 It means that one more pressure exists
 as the average momentum transfer per unit volume:
\begin{equation}
\label{eq=10}
  P_2 = {{\rho_{_{M}}}\over{2}}\biggl(
                                      {{{\vec{\mathit J}}}\over{\rho_{{M}}}}
                               \biggr)^2
 = {{D^2}\over{2}}{{(\nabla\rho_{{M}})^2}\over{\rho_{{M}}}}.
\end{equation}
 One can see that, the sum of the two pressures, $P_1 + P_2$, divided by $\rho$ (we remark that $\rho_{{M}} = m_{}\rho$, see Eq.~(\ref{eq=8}))  reduces to the quantum potential~\cite{Sbitnev2016b}
\begin{equation}
\label{eq=11}
  Q =  {{P_{2}+P_{1}}\over{\rho}}
     =m {{D^2}\over{2}}\biggl(
                                   {{\nabla\rho}\over{\rho}} 
                                   \biggr)^2
 -m D^{2}{{\nabla^2\rho}\over{\rho}} 
 = -2mD^2{{\nabla^{2} R}\over{R}}.
\end{equation}
 $R=\sqrt{\rho}$ is the amplitude distribution of the carriers of $m$ in the volume $\Delta V$.

 Now we need to define the diffusion coefficient $D$ in terms related to the quantum mechanics problems.
 These problems were considered by E. Nelson in his monographs~\cite{Nelson1967,Nelson1985}.
As follows from his article~\cite{Nelson1966} Brownian motion of a carrier of the single mass in the quantum ether is described by the Wiener process with the diffusion coefficient  equal to 
\begin{equation}
    D =  {{\hbar}\over{2m}}.
\label{eq=12}
\end{equation} 

\subsection{\label{sec22}Getting the \Schrodinger-like equation} 

Now dividing Eq.~(\ref{eq=3}) by $\rho$ we may rewrite it in the following view
\begin{equation}
 m \biggl(
 {{\partial {\vec {\mathit v}}}\over{\partial\,t}}
 + ({\vec {\mathit v}}\cdot\nabla){\vec {\mathit v}}
       \biggr) 
  =     -\nabla U({\vec r},t) 
   \;-\;
     \nabla Q
 \; +\; m\nu(t)\,\nabla^{\,2}{\vec {\mathit v}} .
\label{eq=13}
\end{equation}
 Here we believe that the external force ${\vec f({\vec r},t)}$ is conservative and we write it as follows
  ${\vec f}({\vec r},t) = {\vec F}({\vec r},t)/{\Delta V}  = -\nabla U({\vec r},t)/{\Delta V}$.
  The factor $\nu(t) = \mu(t)/\rho_{M}$ is the kinetic viscosity koefficient. It has dimension [length$^2$/time], 
  the same as the diffusion coefficient $D$.
  
  Motion of the carriers that transfer the elementary masses $m$ in the superfluid medium in non-relativistic limit is described by  the modified Navier-Stokes equation~(\ref{eq=13}) + the continuity equation~(\ref{eq=6}). These two equations for real-valued functions, ${\vec{\mathit v}}({\vec r},t)$ and $\rho({\vec r},t)$,  can be reduced to a single equation describing the evolution of a complex-valued function - the \Schrodinger-like equation. But first one should mention that the fundamental theorem of vector calculus states that any vector field can be expressed through sum of irrotational and solenoidal fields. The current velocity  ${\vec{\mathit v}}({\vec r},t)$  can be represented consisting of two components - irrotational and solenoidal vector functions~\cite{KunduCohen2002}:
\begin{equation}
\label{eq=14}
  {\vec{\mathit v}} = {\vec{\mathit v}}_{_{S}} + {\vec{\mathit v}}_{_{R}},
\end{equation}
 where subscripts $S$ and $R$ point to existence of scalar 
  and vector (rotational) potentials  underlying the emergence of these  velocities
 which relate to vortex-free
 and vortex motions of the fluid medium, respectively. 
 They satisfy the following equations
\begin{equation}
\label{eq=15}
 \left\{
    \matrix{
           (\nabla\cdot{\vec{\mathit v}}_{_{S}}) \ne 0, & [\nabla\times{\vec{\mathit v}}_{_{S}}]=0, \cr
           (\nabla\cdot{\vec{\mathit v}}_{_{R}})  =  0, &\, [\nabla\times{\vec{\mathit v}}_{_{R}}]={\vec\omega}. \cr
           }
 \right.
\end{equation}
 Also we shall write down the term $({\vec{\mathit v}}\cdot\nabla){\vec{\mathit v}}$ in Eq. (\ref{eq=13})  in detail
\begin{equation}
\label{eq=16}
  ({\vec{\mathit v}}\cdot\nabla){\vec{\mathit v}} =
 \nabla {\mathit v}^2 /2   +   [{\vec\omega}\times{\vec{\mathit v}}].
\end{equation}
 Here ${\vec\omega} = [\nabla\times{\vec{\mathit v}}]$ is called the vorticity.
   
    Let us now designate the scalar field by a scalar function $S$. Its name is the action.
 Then the irrotational velocity ${\vec{\mathit v}}_{_{S}}$ of the sub-particle is defined as $\nabla S/m$.
 Next, the sub-particle momentum and its kinetic energy   have the following representations
\begin{equation}
\label{eq=17}
 \left\{\,
    \matrix{
           {\vec{\mathit p}} = m{\vec{\mathit v}} = \nabla S + m{\vec{\mathit v}}_{_{R}}, \cr\cr
           {\displaystyle
            m{{{\mathit v}^2}\over{2}} = 
           {{1}\over{2m}}(\nabla S)^2 + m {{{\mathit v}_{_{R}}^2}\over{2}}}. \cr
           }
 \right.
\end{equation}
 As seen in the bottom line, the kinetic energy of the fluid motion is equal to sum of the kinetic energies of irrotational and solenoidal motions.
  Now taking into account the expression~(\ref{eq=16})
 we may rewrite the Navier-Stokes equation~(\ref{eq=13}) in a more detailed form by regrouping its terms~\cite{Sbitnev2016b}
\begin{eqnarray}
\nonumber
&&
 \nabla\biggl(
  \underbrace {
  {{\partial~}\over{\partial\,t}}S + {{1}\over{2m}}(\nabla S)^2 + {{m}\over{2}}{\mathit v}_{_{R}}^2
              + U + Q - \nu(t)\nabla^{2}S}_{f_1({\vec r},t)}
       \biggr)
\\
&&
 =  \underbrace {
 -m{{\partial~}\over{\partial\,t}}{\vec{\mathit v}}_{_{R}}
   - m [\,{\vec\omega}\times{\vec{\mathit v}}_{_{}}\,]
    +\nu(t)m\nabla^2 {\vec{\mathit v}}_{_{R}}}_{{\vec f}_2({\vec r},t)}.
\label{eq=18}
\end{eqnarray}
 Here the brackets single out two qualitatively different functions, scalar-valued fuction $f_1({\vec r}, t)$ and vector-valued function 
${\vec f}_2({\vec r}, t)$. Depending on application of the $\nabla$-operator, either as the inner product or the cross product, we may extract either the left side or the right side of this expression.

 One can see that the scalar-valued function $f_1({\vec r}, t)$ with omitted the last two terms represents the Hamilton-Jacobi equation. In pair with the continuity equation they describe motion of a deformable medium populated by the carriers having the elementary mass $m$. However the Hamilton-Jacobi equation loaded by the last two terms describes the motion of the quantum medium subjected to zero-point fluctuations because of fluctuating the kinetic viscosity coefficient $\nu(t)$. These fluctuations describe exchange of the energy within the superfluid quantum space (SQS)~\cite{SbitnevFedi2017}.
 
 Now by applying the $\nabla$-operator to Eq.~(\ref{eq=18}) as the scalar product one can see that
  $(\nabla\cdot{\vec f}_2({\vec r}, t))$ vanishes. So, the function $f_1({\vec r}, t)$ should be constant
\begin{equation}
\label{eq=19}
    {{\partial~}\over{\partial\,t}}S + {{1}\over{2m}}(\nabla S)^2 + {{m}\over{2}}{\mathit v}_{_{R}}^2
      + U 
      + Q
      + \underbrace{\nu(t)m f(\rho)}
      = C_0.
\end{equation}
 Here $C_0$ is the integration constant.
  We rewrote the term $\nu(t)\nabla^{2}S$ 
  as $\nu(t)m f(\rho)$ by taking into  account that
\begin{equation}
  \nabla^{2}S = m(\nabla{\vec{\mathit v}})
   = - m{{d\,}\over{d\,t}}\ln(\rho)
   = - m f(\rho)
\label{eq=20}
\end{equation}
 and the term ${{d\ln(\rho)}/{d\,t}}$ comes from the continuity equation
\begin{equation}
     {{\partial\,\rho_{_{}}}\over{\partial\,t}} +(\nabla{\vec{\mathit v}})\rho_{_{}}
 =  {{d\,\rho_{_{}}}\over{d\,t}}
     + \rho_{_{}}(\nabla{\vec{\mathit v}})
 = 0
 \;\rightarrow\;
 {{d\ln(\rho)}\over{d\,t}} + (\nabla{\vec{\mathit v}}) =0.
\label{eq=21}
\end{equation}
 Now we can use the wave function in the polar representation 
\begin{equation}
\label{eq=22}
  \Psi = \sqrt{\,\rho\,}\exp\{{\bf i}S/\hbar\}
\end{equation}
  in order to get the \Schrodinger-like equation
\begin{equation}
\label{eq=23}
\hspace{-8pt}
  {\bf i}\hbar\,{{\partial\Psi}\over{\partial\,t}}=
  {{1}\over{2m}}(-{\bf i}\hbar\nabla + m{\vec{\mathit v}}_{_{R}})^2\Psi
      + \underbrace{\nu(t)m f\bigl(|\Psi|^{2}\bigr)}\Psi 
      + U({\vec r})\Psi  
       - C_0\Psi
\end{equation}
 
One can see that this equation is almost like the Gross-Pitaevski equation~\cite{AbidEtAl2003,RobertsBerloff2001} for exception of two moments: (a) the term $\rho=|\Psi|^2$ enters in this equation through the full derivative from logarithm by time; (b) the viscosity coefficient, $\nu(t)$, is the fluctuating variable of time, at that the average on time vanishes and its variance is not-zero, see Eq.~(\ref{eq=5}). It can mean that the above equation is the Langevin equation with an non-linear noise source. From here we can conclude that the wave function being a solution of this equation may manifest small fluctuations similar to  the Brownian  ripples.

\subsection{\label{sec23}Computations of the Bohmian trajectories} 

Here we will follow the computation methods stated in~\cite{OriolsMompart2013}. Two equations, the continuity equation~(\ref{eq=6}) and the quantum Hamilton-Jacobi equation~(\ref{eq=19}) are used for computing the Bohmian trajectories.  For the sake of simplicity we omit in Eq.~(\ref{eq=19}) the integration constant $C_0$ and the stochastic fluctuation term $\nu(t)mf(\rho)$. According to~\cite{OriolsMompart2013} we replace  in the continuity equation the mass density distribution $\rho_{M}({\vec r},t)$ by a function $C_{\rho}({\vec r, t})$ according to the following relationship $\rho_{M}=m\rho=m\exp\{2C_{\rho}\}$. We get the following equations:
\begin{eqnarray}
\label{eq=24}
&&
  {{\partial }\over{\partial\,t}} C_{\rho}
  + {{1}\over{2m}}\Bigl(
                                     \nabla^2 S + 2(\nabla S\!\cdot\!\nabla C_{\rho})
                            \Bigr)
                            + ({\vec{\mathit v}}_{_{R}}\!\cdot\!\nabla C_{\rho})=0,\\
\nonumber \\                            
&&
    {{\partial~}\over{\partial\,t}}S + {{1}\over{2m}}(\nabla S)^2 \
- {{\hbar}\over{2m}}\Bigl(
                                       \nabla^2C_{\rho} + (\nabla C_{\rho})^2
                               \Bigr)
      + U
      + {{m}\over{2}}{\mathit v}_{_{R}}^2 
      = 0.
\label{eq=25}                            
\end{eqnarray}
These equations differ from those shown in~\cite{OriolsMompart2013}
by the presence of the solenoidal velocity ${\vec{\mathit v}}_{_R}$ that reflects the existence of the rotational motions. If the vorticities are absent (that is, at discarded terms containing ${\vec{\mathit v}}_{_R}$) then the Bohmian trajectories are computed by the algorithms given in~\cite{OriolsMompart2013}. 
\begin{figure}[htb!]
  \begin{picture}(200,135)(0,0)
      \includegraphics[scale=1, angle=0]{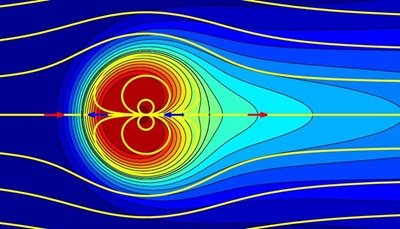}
  \end{picture}
  \caption{
   Steady vortex avenue confined by transverse flow with a dipole source inside 
   and a uniform background flow outside, taken from~\cite{BazantMoffatt2005}.
   Yellow streamlines outside of the vortex area mean possible Bohmian trajectories.
   }
  \label{fig=1}
\end{figure}

Note that the Bohmian trajectories extracted from the solution of these equations and covering  everywhere the physical scene under consideration are geodesic trajectories showing possible optimal paths for particles. Since particles consist in bundles of swirling flows, they disturb the surrounding Bohmian trajectories, in such a way that these trajectories flow around the vortex region, see Fig.~\ref{fig=1}. The picture has been constructed with attracting the canonical problem of steady advection-diffusion around an absorbing, circular cylinder in a uniform background flow in 2D geometry~\cite{BazantMoffatt2005}.

In order to see such a picture we need to solve an additional equation marked in Eq.~(\ref{eq=18}) by the bracket $f_2({\vec r},t)$. For extracting this equation we should apply  to~(\ref{eq=18}) the curl operator. Observe that in this case we go to a local coordinate system with the origin at the center of the vortex and which slides along an allocated Bohmian trajectory at the velocity of ${\vec{\mathit v}}$. 
One can see, that this picture demonstrates de Broglie's representation of the pilot-wave guiding the particle (a vortex clot) along the optimal trajectory~\cite{deBroglie1960,deBroglie1987}.

\section{\label{sec3}Motion of vortices along the Bohmian trajectories}

By multiplying Eq.~(\ref{eq=18}) by the curl operator from the left
we find that the first term vanishes and we obtain the equation for the vorticity
\begin{equation}
 {{\partial\, {\vec\omega}}\over{\partial\,t}}
 + ({\vec\omega}\cdot\nabla){\vec{\mathit v}}
 = \nu(\,t\,)\nabla^{2}{\vec\omega}.
\label{eq=26}
\end{equation}
 The rightmost term describes dissipation of the
energy stored in the vortex. As a result, the vortex should disappear in time.
By omitting this rightmost term (that is, assuming $\nu=0$) we open the action to the Helmholtz theorem~\cite{Lighthill1986}: (i)~if fluid particles form, in any moment, a vortex line, then the same particles support it both in the past and in the future; (ii)~an ensemble of vortex lines traced through a closed contour forms a vortex tube; (iii)~the intensity of the vortex tube is constant along its length and does not change in time. The vortex tube either (a)~goes to infinity by both endings; or (b)~these endings lean on boundary walls containing the fluid; or (c)~these endings are locked to each on other forming a vortex ring.

At consideration of the vortices evolving along the Bohmian trajectory it is convenient to choose a local coordinate system, the origin of which is linked with the moving particle. In this case the vortex ideally simulates the particle. 
Spherical and toroidal coordinate systems are perfect for this purpose in 3D geometry.
In order to show some specific features of vortices in their cross-section, it is usually employed the cylindrical coordinate system in 2D geometry.
 Here we consider the main, topological features of the vortex. Let us look on the vortex tube in its cross-section which is oriented along the $z$-axis and its center is placed in the coordinate origin of the plane $(x, y)$.
  Eq.~(\ref{eq=26}), written down for the cross-section of the vortex, looks as follows~\cite{Sbitnev2016c}
\begin{equation}
  {{\partial\, {\omega}}\over{\partial\,t}} =
 \nu(\,t\,)\Biggl(
    {{\partial^{\,2}\omega}\over{\partial\,r^{2}}}
 +{{1}\over{r}}{{\partial\,\omega}\over{\partial\,r}}
               \Biggr).
\label{eq=27}
\end{equation}
 A general solution of this equation has the following view
\begin{equation}
\label{eq=28}
 \omega(r,t)={{\mit\Gamma}\over{4\Sigma(\nu,t,\sigma)}}
  \exp
  \matrix{
  \left\{\displaystyle
      -{{r^2}\over{4\Sigma(\nu,t,\sigma)}}  
      \right\}},
\end{equation}
\begin{equation}
\label{eq=29}
\hspace{-8pt}
 {\mathit {v}}(r,t)
= {{1}\over{r}}\int\limits_{0}^{r}\omega(r',t)r'dr'
={{\mit\Gamma}\over{2 r}}
 \matrix{
             \left( 1 -
  \exp
  \left\{\displaystyle
      -{{r^2}\over{4\Sigma(\nu,t,\sigma)}}  
      \right\}
           \right).
          }
\end{equation}
 Here $\mit\Gamma$ is the integration constant having dimension [length$^2$/time]
 and the denominator ${\mit\Sigma}(\nu,t,\sigma)$ has a view
\begin{equation}
   {\mit\Sigma}(\nu,t,\sigma) =
   \int\limits_{0}^{t} \nu(\tau) d\tau + \sigma^{2}.
\label{eq=30}
\end{equation} 
 Here $\sigma$ is an arbitrary constant such that the denominator is always positive.
 
 Since the time average of the viscosity is zero, see Eq.~(\ref{eq=5}) , the above solutions,~(\ref{eq=28}) and~(\ref{eq=29}), at large times, $t\gg1$, reduce to
\begin{eqnarray}
 \omega(r) & =& {{\mit\Gamma}\over{4\sigma^2}}
  \exp
  \matrix{
  \left\{\displaystyle
      -{{r^2}\over{4\sigma^2}}  
      \right\}},
\label{eq=31}
\\
\label{eq=32}
\hspace{-8pt}
 {\mathit {v}}(r)
&=&{{\mit\Gamma}\over{2 r}}
 \matrix{
             \left( 1 -
  \exp
  \left\{\displaystyle
      -{{r^2}\over{4\sigma^2}}  
      \right\}
           \right).
          }
\end{eqnarray}
This solution is named the Gaussian coherent vortex cloud~\cite{ProvenzaleEtAl2008,NegrettiBillant2013,KleinebergFriedrich2013} that is permanent in time.
\begin{figure}[htb!]
  \begin{picture}(200,90)(5,10)
      \includegraphics[scale=0.5, angle=0]{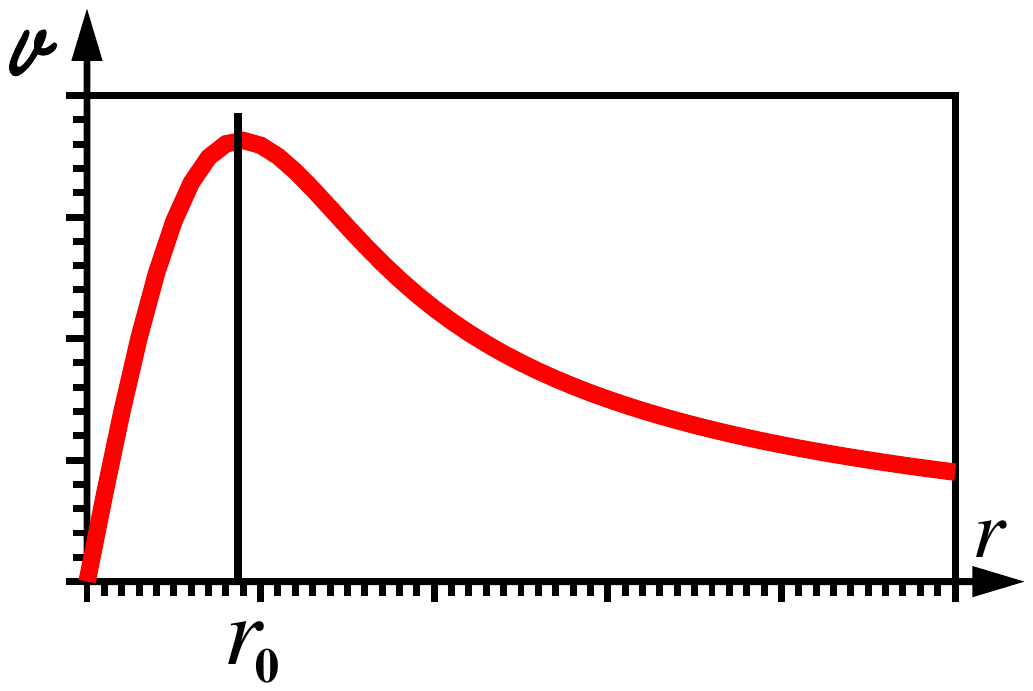}
  \end{picture}
  \caption{
   Orbital velocity ${\mathit v}(r)$ vs. $r$. Point $r_0$ marks the boundary of the vortex core.
   }
  \label{fig=2}
\end{figure}
The orbital velocity~(\ref{eq=32}) as a function of $r$ is shown in Fig.~\ref{fig=2}. One can see that the velocity grows from zero to some maximal value as $r$ increases from zero to $r_0$. Further  the velocity begins to decrease monotonically up to zero at $r$ tending to infinity. The point $r_0$ is the boundary of the vortex core. 
 Further we adopt the vortex core as a visit-card of the vortex, remembering that eddy fluxes are maximal on the wall of the core and monotonically decrease as the distance from the core. From this point of view, the vortex in the cylindrical coordinate system, for example, looks like a tube with the radius $a = r_0$ ends of which go to infinity.
  It permits to consider any topological transformations of the vortex without loss of generality. 
  
\subsection{\label{sec31}Toroidal vortex}  
  
\begin{figure}[htb!]
  \begin{picture}(200,110)(0,3)
      \includegraphics[scale=0.25, angle=0]{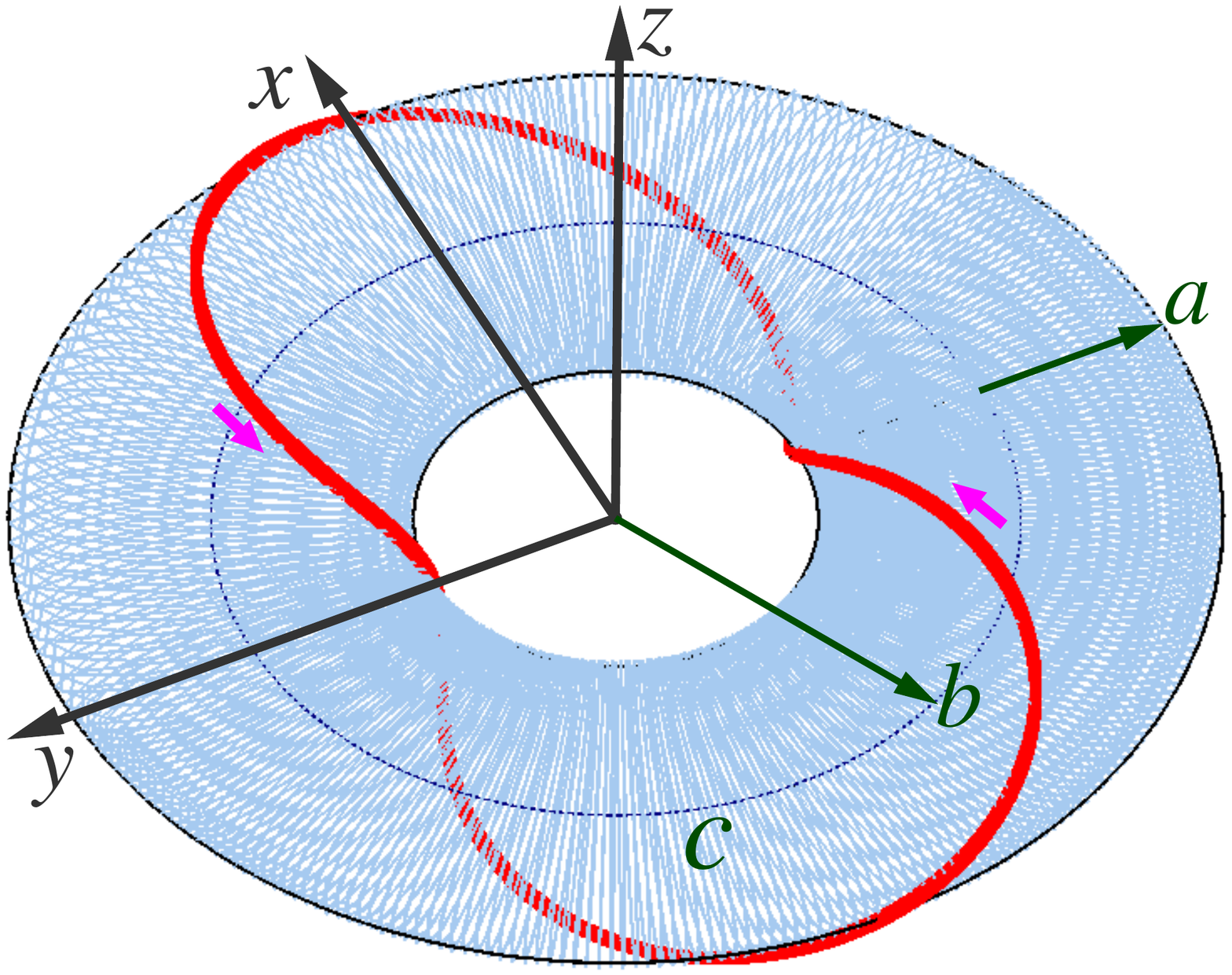}
  \end{picture}
  \caption{
   Toroidal vortex.
   Radius of the tube $a=2$ and radius of the torus $b=4$. The center line of the perimeter of the tube is indicated by the letter $c$.
Pink arrows point out motion along the helicoidal ring drawn by red color.
   }
  \label{fig=3}
\end{figure}

Let us glue the opposite ends of the tube according to the  second Helmholtz's theorem. 
As a result we get the toroidal vortex with radius of the torus $b$ as shown in Fig.~\ref{fig=3}. 
It is a standard topological procedure obtaining the torus from the tube by means of the evolution transformations: vortex tube $\rightarrow$ vortex torus $\rightarrow$ and etc.~\cite{Sonin2012,SbitnevFedi2017}. The torus obtained lies in the (x,y) plane, and the z-axis is perpendicular to it.
Position of points on this torus in the Cartesian coordinate system is given by the following set of equations
\begin{equation}
\label{eq=33}
\hspace{-14pt}
\left\{
\matrix{
  x=(b_{}+a_{}\cos(\omega_{\,0}t+\phi_{\,0}))\cos(\omega_{1}t+\phi_{1}),\cr
  y=(b_{}+a_{}\cos(\omega_{\,0}t+\phi_{\,0}))\sin(\omega_{1}t+\phi_{1}),\cr
 ~z= a_{}\sin(\omega_{\,0}t+\phi_{\,0}).\hspace{90pt}\cr
       }
\right.
\end{equation}
The frequency $\omega_{\,0}$ is that of the rotation of tube points about the central ring of the tube marked by the letter c in Fig.~\ref{fig=2}. The frequency $\omega_{1}$ is that of rotation about the $z$ axis.
 Parameter $a_{}$ is the radius of the tube. And $b_{}$ is the radius of the torus - the distance from the origin to the circle pointed by  c in Fig.~\ref{fig=3}. The phases $\phi_{\,0}$ and $\phi_{1}$ can have arbitrary values between  0 to $2\pi$. 

Depending on choosing the frequencies  $\omega_{\,0}$ and  $\omega_{\,1}$  a helicoidal ring can be spooled $n$ times either around the tube ($n$ is represented by the ratio $\omega_{\,0}$ to $\omega_{\,1}$) or around $z$ axis ($n$ is represented by the ratio $\omega_{\,1}$ to $\omega_{\,0}$). The helicoidal ring shown in Fig.~\ref{fig=3} in red is drawn at choosing $\omega_{\,1} = \omega_{\,0}/2$  and at $\phi_{\,0}=\phi_{\,1}=0$. In this case, a point moving along the helicoidal ring performs two turns about the tube. In general, the phases, $\phi_{\,0}$ and $\phi_{\,1}$ can have arbitrary values within the interval $[0,2\pi)$. It means that the surface area of the torus can be filled by set of  the helicoidal rings everywhere densely  at shifting the phases on infinitesimal increments.
    
\subsubsection{\label{sec311}Topological transformations}    

Let us begin from recalling the volume and the surface area of a torus 
\begin{eqnarray}
\label{eq=34}
   V &=&  2\pi^2 ba^2, \\
   S &=& 4\pi^2 ba.
\label{eq=35}   
\end{eqnarray}
 Here $a$ is the radius of the tube representing a body of the torus and $b$ is the radius of the circumference that is the central axis of the tube, see Fig.~\ref{fig=3}.
 
Observe that the volume and the surface area vanish as the radius $a$ goes to zero. In this case the torus degenerates to an one-dimensional ring that belongs to a class of one-dimensional objects called strings in the string theory~\cite{t_Hooft2014}.
\begin{figure}[htb!]
  \begin{picture}(200,360)(0,3)
      \includegraphics[scale=0.6, angle=0]{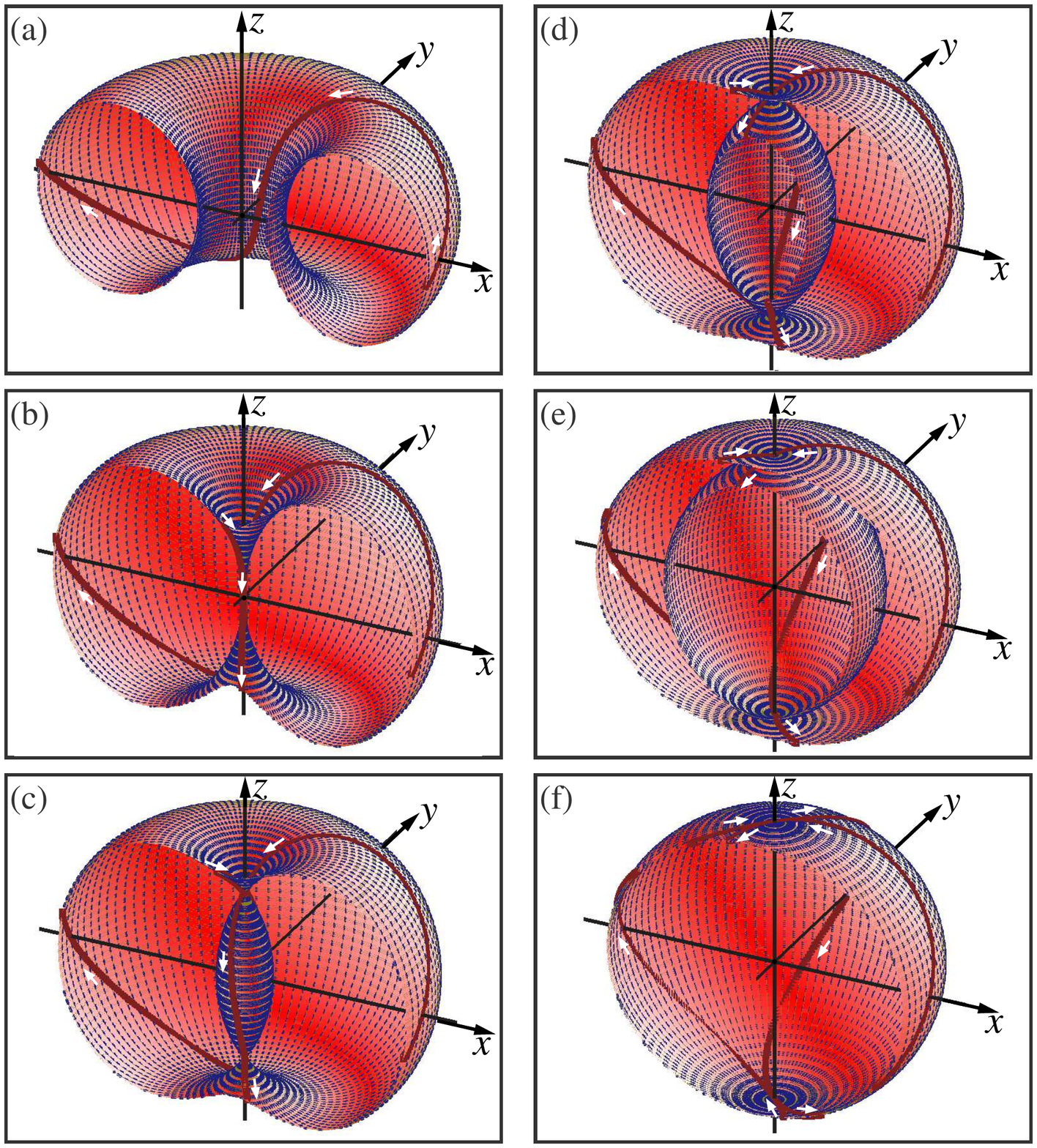}
  \end{picture}
  \caption{
    Topological transformation of the torus to the spindle torus from (a) to (f). The radius of the tube is fixed $a=2$. The transformation is made through variation of the radius of the circumference {\it c}: (a)~$b=3$; (b)~$b=2$; (c)~$b=1.5$; (d)~$b=1$; (e)~$b=0.5$; (f)~$b=0.01$. Dark red strips show transformations of the helicoidal ring drawn on the previous figure. White arrows indicate the motion around the helicoidal ring.
   }
  \label{fig=4}
\end{figure}

\begin{figure}[htb!]
  \begin{picture}(220,120)(0,0)
      \includegraphics[scale=0.25]{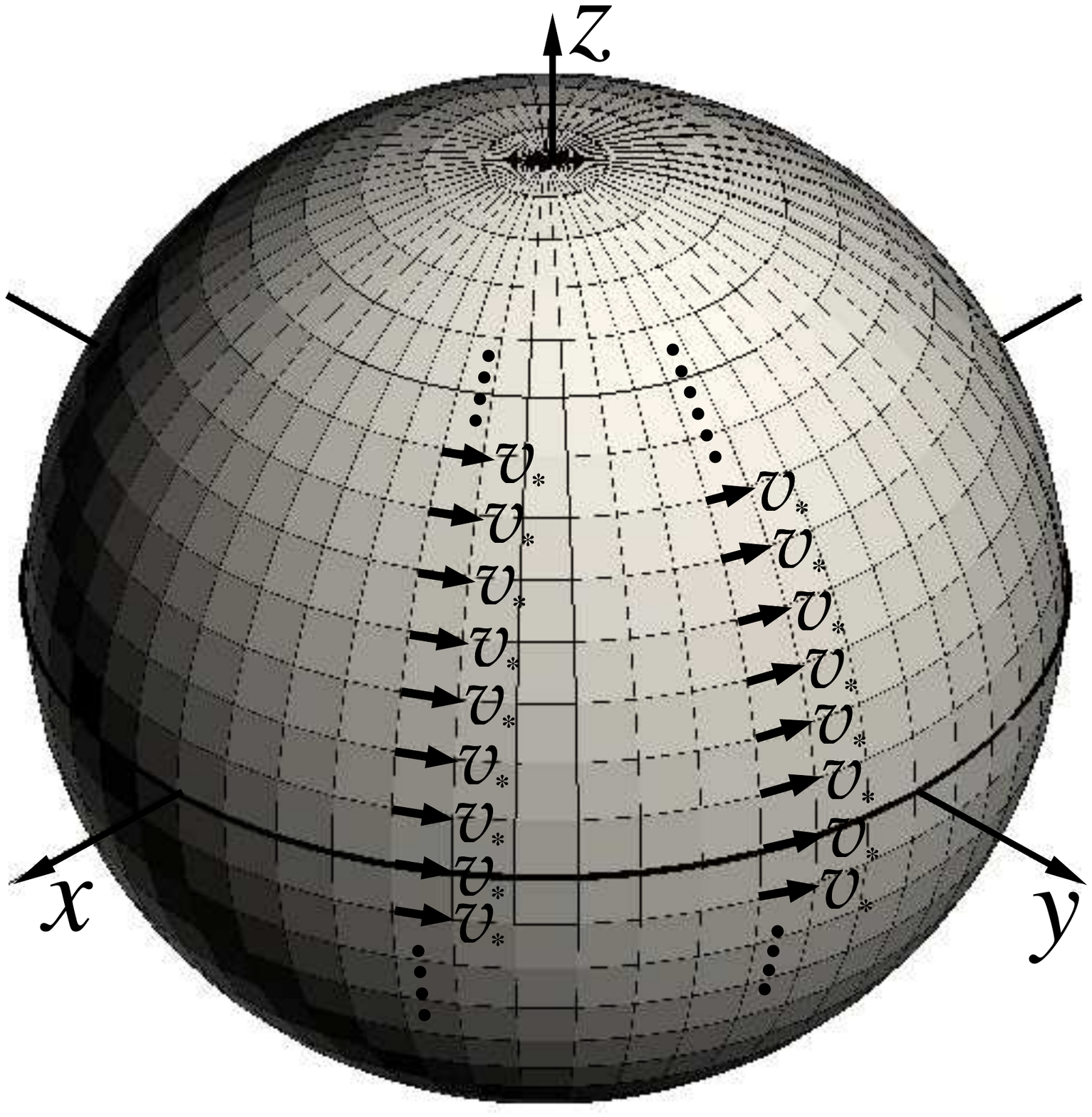}
  \end{picture}
  \caption{
 Qualitative presentation of the self-organizing flow on the wall of the vortex core~\cite{Sbitnev2016c}:  $a_{0}=1$, $b_{1}\approx 0$: the orbital velocity ${\vec\mathit v}_{*}=\vec{\mathit v}_{+}+{\vec\mathit v}_{-}=(2{\mathit v}_{x}, 2{\mathit v}_{y}, 0)$:
  longitudinal components of the velocities coincide,  transversal components have opposite orientations.
  }
  \label{fig=5}
\end{figure}

There is another possibility when volume and the surface area of the torus go to zero under the radius $a$ being fixed. In this case the radius $b$ should tend to zero. When $b$  becomes smaller then $a$ the axis of rotation intersects the tube. Along the axis of rotation we can see emergence of the central "spindle", see Fig.~\ref{fig=4}. As $b$ tends to zero the central radius of the spindle grows until it reaches the radius $a$, as shown in Figs.~\ref{fig=4}(c) to~\ref{fig=4}(f).  
In the last case one can see emergence of a sphere of the radius $a$.
At $b=0$ it seems the vortex ball disappears. In this case it transforms to the rotating sphere, Fig.~\ref{fig=5}.
Observe, however, that at $b$ small but no equal to zero the surface of the vortex ball  represents a double layer with opposite orientation of the unit normal vectors.

\begin{figure}[htb!]
  \begin{picture}(200,70)(0,3)
      \includegraphics[scale=0.5, angle=0]{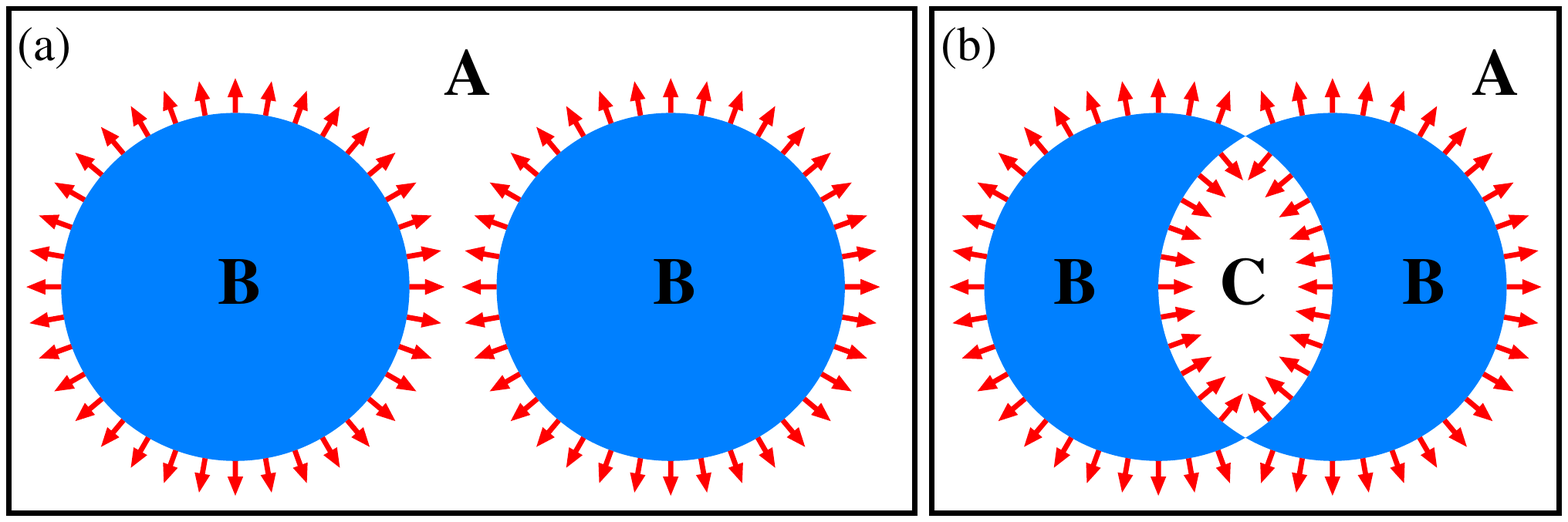}
  \end{picture}
  \caption{
  Cross-section of the torus along the medial plane $(x,z)$, $y=0$. (a) there are two separate regions, outer region {\bf A} and inner region {\bf B}, when $b\ge a$; (b) when $b < a$ extra region, named the spindle region {\bf C}, arises. Red arrows mark orientations of the unit normal vectors of the tube surface.
   }
  \label{fig=6}
\end{figure}

In order to trace this transformation let us look at the convergence of two circles in the cross-section of the torus as the distance between their centers decreases, see Fig.~\ref{fig=6}. One can see that two separate regions of the torus, the outer region {\bf A} and inner region {\bf B}, exist until the spindle arises. The inner region of the spindle, the region {\bf C}, arising when $b>a$ is an extension of the outer region {\bf A}. 
So, the inner region of the torus, the region of the tube, diminishes with decreasing $b$ and it disappears when $b$ vanishes. 
In Fig.~\ref{fig=6} red arrows  mark orientations of the unit normal vectors of the surface area of the tube.  Observe that there is an unbroken continuity in the orientations of the vectors when shifting along the surface even on borders of crossing surfaces of different volumes, as shown in Fig.~\ref{fig=6}(b). Therefore, when the radius $b$ reaches zero and the volume of the tube vanishes, the surface of the sphere, visible in Fig.~\ref{fig=4}(f), acquires a double coating, each of which has oppositely oriented unit normal vectors. 
This double folded surface covers the core vortex, where the orbital velocity reaches maximal magnitudes. 
So, there are two orbital velocities which have opposite oriented  velocity components along the $z$-axis  and the longitudinal components coincide~\cite{Sbitnev2016c}. Addition of the velocities gives the total rate of rotation about the $z$-axis, Fig.~\ref{fig=5}.

\begin{figure}[htb!]
  \begin{picture}(200,120)(0,10)
      \includegraphics[scale=0.6, angle=0]{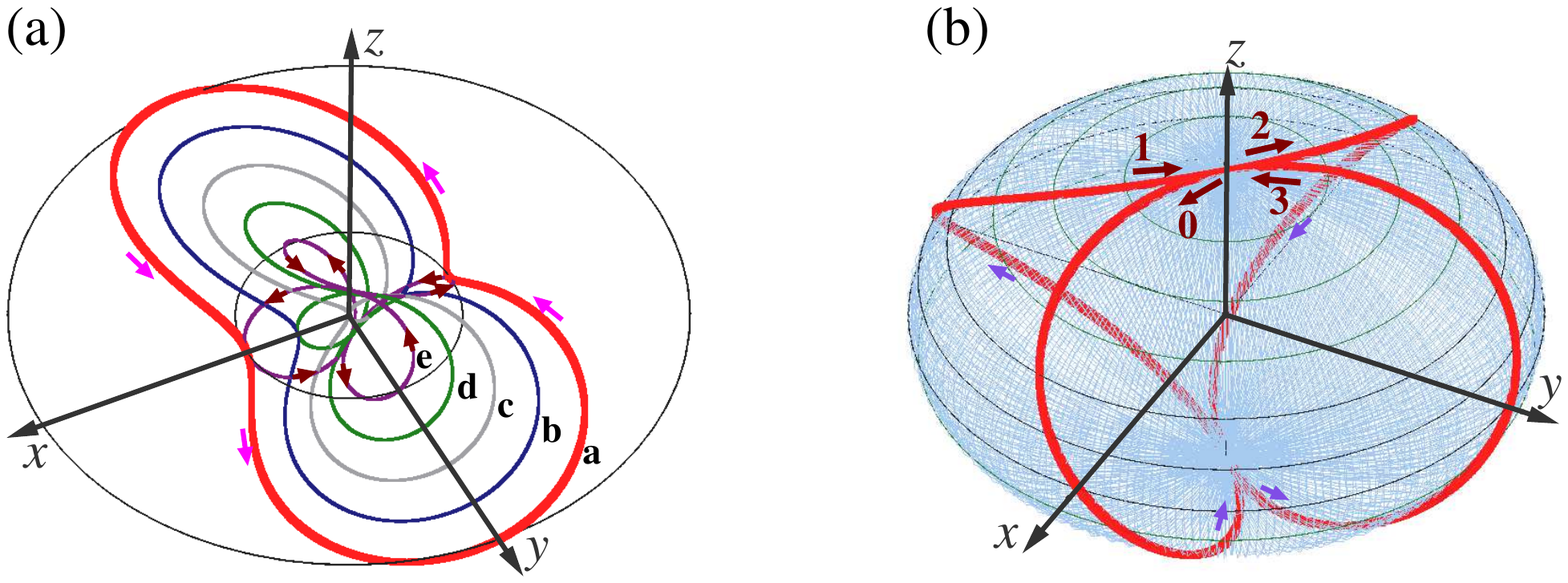}
  \end{picture}
  \caption{
  Topological transformation of the helicoidal ring shown in Fig.~\ref{fig=3}  in red to the ring on the sphere with the double coating: (a) sequential transformation: {\bf  a})  $b=4$, {\bf b})  $b=3$, {\bf c})  $b=2$, {\bf d}) $b=1$, {\bf e}) $b=0.001$; 
  (b)  the ring {\bf  e} is shown on the torus transformed into the double coated sphere having the parameters $a=2$, $b=0.001$.
Arrows near red helicoidal ring show direction of  the bypass of a test point along the ring.
      }
  \label{fig=7}
\end{figure}
The helicoidal ring shown in Fig.~\ref{fig=3} in red when subjected to the topological transformation as the radius $b$ decreases from 4 to 0.001 is shown in Fig.~\ref{fig=7}(a). One can see formation of a double loop (the transformed helicoidal ring) shown in Fig.~\ref{fig=7}(b) in red. Here the direction of the bypass along the loop is shown by arrows drawn nearby the ring. 
One can see that the return of the test point to an initial state (the top point in the $z$-axis) occurs after the second bypass of this point. After the first revolution the arrow 0 changes its orientation to opposite. It is the arrow $1\!\rightarrow\!2$. The arrow 2 after the next revolution comes to the position 3. This arrow has the same initial orientation as the arrow 0. So, the full revolution is $2\cdot360^\circ = 720^\circ$. Conditionally speaking, such an alternation of the orientations at each revolution on $360^\circ$ can be expressed by the alternation of opposite colors placed on the color wheel, say, green (spin up), magenta (spin down), green (spin up), magenta (spin down), etc. 
As for the vortex ball shown in Fig.~\ref{fig=5} one can imagine that the ball undergoes color changing each revolution on $360^\circ$.
Observe that this revolution  shows a good agreement with  rotation of a 1/2--spin particle which can be described by the special unitary group {SU(2)}.   
Further we will return to this group at studying behavior of the 1/2--spin in a magnetic field. Now we summarize the results.   
        
\section{\label{sec4}Conclusion}        

The Nobel laureate in Physics 2003 Vitaly Ginzburg in his Nobel speech~\cite{Ginzburg2007a} highlighted a number of the great issues which will stand before humanity in the near future. Among them he mentioned interpretation of the non-relativistic quantum mechanics. Content of this article directly relates to the solution of this problem.  
The article rehabilitates the early idea of Louis de Broglie on the pilot-wave~\cite{deBroglie1960,deBroglie1987}, accompanying a particle from the moment of its creation on a source until it will be registered by a detector. 

According to the de Broglie ideas the wave function $\psi$ normed on unit has a statistical significance in the usual quantum mechanical formalism. However it has its own replica, a physical wave ${\mathit v}$, which is connected with the mathematical wave $\psi$ by the relation $\psi = C{\mathit v}$, where $C$ is a normalizing factor. De Broglie writes~\cite{deBroglie1987}: "the $\psi$ wave has the nature of a subjective probability representation formulated by means of the objective ${\mathit v}$ wave." In other words, the ${\mathit v}$ wave is a really existing wave whereas the $\psi$ wave is a mathematical copy of the physical wave ${\mathit v}$. And further he writes~\cite{deBroglie1987}: " for me, the particle, precisely located in space at every instant, forms on the ${\mathit v}$ wave a small region of high energy concentration, which may be likened in a first  approximation, to a moving singularity."

The results of the article say that the interpretation given by de Broglie~\cite{Smyk2011a}
along with the expanded version given by Bohm~\cite{Bohm1952a,Bohm1952b}
is the most correct.  To confirm this assertion it makes sense to begin with the idea of the an ether pervading all the physical space, but now on a fundamentally new basis. Now we can call this medium a superfluid quantum space (SQS). 
The behavior of this vast ocean of energy can be described in the non-relativistic limit with involvement of the hydrodynamical Navier-Stokes equation slightly modified with the aim to describe a quantum superfluid nature of this medium. 
This equation has a vast class of solutions that can contain those showing fields of irrotational velocities and the existence of vortex solutions.

The first class of solutions gives the velocities field stemming from gradient of the scalar function $S$, named the action.  The irrotational velocity ${\vec{\mathit v}} \sim \nabla S$. In pair with solutions of the continuity equation one can combine the complex-valued function 
   $\psi({\vec r},t) = \sqrt{\rho({\vec r},t)} \exp\{ {\bf i}S({\vec r},t)/\hbar \}$
which describes a wavy behavior of the SQS. This polar form of the wave function has been shown by de Broglie in~\cite{deBroglie1987} in connection with describing the ${\mathit v}$ wave.

The de Broglie's ${\mathit v}$ wave is the pilot-wave guiding the particle along an optimal path, that is the Bohmian trajectory, from its creation up to detection. Remarkably, this function bears a wave pattern repeating in its "wavy memory" all the peculiarities of a physical scene where the observed events unfold. Due to this fact a particle moving through SQS receives all information about the environment. This mechanism is as follows~\cite{FeynmanHibbs1965}: the particle perturbs a sensitive SQS and the waves spreading from the particle are reflected by the obstacles present in the surrounding space and by returning to the particle create an interference pattern in the vicinity of its location~\cite{Sbitnev2013a}. 
This constructive and destructive interference pattern provides a path for the particle further.
It is instructive to note here, that an alike information mechanism for a droplet bouncing on a fluid surface subjected by a supercritical Faraday oscillations was described by Couder and Fort~\cite{CouderForte2006} and the right interpretation in light of the de Broglie's pilot-wave was given by Bush~\cite{Bush2015a}.

As for the particle, de Broglie wrote~\cite{deBroglie1987} that it occupies "a small region of high energy concentration, which may be likened in a first approximation, to a moving singularity."
The particle representing a small region of high energy concentration, is simulated as a vortex clot stemming from the second solution of the Navier-Stokes equation.
The coordinate origin of the local coordinate system, within which the vortex clot is represented, moves along the Bohmian trajectory with the velocity ${\vec{\mathit v}} =\nabla S/m$.
Obviously, the vortex clot deforms the Bohmian trajectories in the vicinity of its own location, so these deformed trajectories looks as  streamlines around the vortex clot, as shown in Fig.~\ref{fig=1}.

Now it can be argued that a complete and self-sufficient interpretation of the quantum mechanics is the de Broglie - Bohm interpretation, upgraded by the fact that the hidden medium is the superfluid quantum space rather than an undefined ether. This space is both a source of waves reproducing  the interference patterns about the environment, and creates vortices which simulate particles.

\begin{acknowledgements}
The author thanks Marco Fedi for useful and valuable remarks and offers.
\end{acknowledgements}

\bibliographystyle{spmpsci}      

%
%

\end{document}